# Data-driven Nucleus Subclassification on Colon H&E using Style-transferred Digital Pathology

**Lucas W. Remedios,[a*] Shunxing Bao,[b] Samuel W. Remedios,[c,d] Ho Hin Lee,[a] Leon Y. Cai,[e] Thomas Li,[e] Ruining Deng,[a] Nancy R. Newlin[a], Adam M. Saunders,[b] Can Cui,[a] Jia Li,[f] Qi Liu,[f,g] Ken S. Lau,[g,h,i] Joseph T. Roland[h], Mary K Washington,[n] Lori A. Coburn,[j,k,l,m] Keith T. Wilson,[j,k,l,m,n] Yuankai Huo,[a,b] Bennett A. Landman,[a,b,e]**

[a]Vanderbilt University, Department of Computer Science, Nashville, USA
[b]Vanderbilt University, Department of Electrical and Computer Engineering, Nashville, USA
[c]Johns Hopkins University, Department of Computer Science, Baltimore, USA
[d]National Institutes of Health, Department of Radiology and Imaging Sciences, Bethesda, USA
[e]Vanderbilt University, Department of Biomedical Engineering, Nashville, USA
[f]Vanderbilt University Medical Center, Department of Biostatistics, Nashville, USA
[g]Vanderbilt University Medical Center, Center for Quantitative Sciences, Nashville, USA
[h]Vanderbilt University Medical Center, Epithelial Biology Center, Nashville, USA
[i]Vanderbilt University School of Medicine, Department of Cell and Developmental Biology, Nashville, USA
[j]Vanderbilt University Medical Center, Division of Gastroenterology, Hepatology, and Nutrition, Department of Medicine, Nashville, USA
[k]Vanderbilt University Medical Center, Vanderbilt Center for Mucosal Inflammation and Cancer, Nashville, USA
[l]Vanderbilt University School of Medicine, Program in Cancer Biology, Nashville, USA
[m]Veterans Affairs Tennessee Valley Healthcare System, Nashville, TN, USA
[n]Vanderbilt University Medical Center, Department of Pathology, Microbiology, and Immunology, Nashville, TN, USA

## Abstract

**Purpose**: Cells are building blocks for human physiology; consequently, understanding the way cells communicate, co-locate, and interrelate is essential to furthering our understanding of how the body functions in both health and disease. Hematoxylin and eosin (H&E) is the standard stain used in histological analysis of tissues. Microscope slides of H&E-stained tissues are widely available in both clinical and research settings. While H&E is ubiquitous and reveals tissue microanatomy, the classification and tracking of cell subtypes often requires expert knowledge and the use of specialized stains, such as immunofluorescence staining. To reduce both the manual annotation burden and reliance on specialized staining technologies, artificial intelligence has been proposed for the automatic classification of cell types on H&E slides. For example, the recent Colon Nucleus Identification and Classification (CoNIC) Challenge focused on labeling 6 cell types on imaging of H&E stains from the human colon. However, this is a very small fraction of the number of potential cell subtypes within the intestines. Specifically, the CoNIC Challenge was unable to classify epithelial subtypes (progenitor, enteroendocrine, goblet), lymphocyte subtypes (B, helper T, cytotoxic T), and connective subtypes (fibroblasts). To approach this problem, we propose to use inter-modality learning to label previously un-labelable cell types on H&E.

**Approach**: We take advantage of the cell classification information inherent in whole slide images (WSIs) of multiplexed immunofluorescence (MxIF) histology to create cell level annotations for 14 subclasses. We performed style transfer on the same MxIF tissues to synthesize realistic virtual H&E which we paired with the MxIF-derived cell subclassification labels. We evaluated the efficacy of using a supervised learning scheme where the input was realistic-quality virtual H&E and the labels were MxIF-derived cell subclasses. We assessed our model on a testing set of virtual H&E from both the ascending colon and terminal ileum, and further evaluated generalization on real H&E using an external set of publicly available multi-site colon data.

**Results**: On virtual H&E, we were able to classify helper T cells and epithelial progenitors with positive predictive values of $0.34 \pm 0.15$ (prevalence $0.03 \pm 0.01$) and $0.47 \pm 0.1$ (prevalence $0.07 \pm 0.02$) respectively, when using ground truth centroid information. On real H&E we could classify helper T cells and epithelial progenitors with upper bound positive predictive values of 0.43 ± 0.03 (parent class prevalence 0.21) and 0.94 ± 0.02 (parent class prevalence 0.49) when using ground truth centroid information.

**Conclusions**: This is the first work to provide cell type classification for helper T and epithelial progenitor nuclei on H&E. The code is available: github.com/MASILab/nucleus_and_cell_classification_on_he

*Lucas W. Remedios, E-mail: lucas.w.remedios@vanderbilt.edu

## 1    Introduction

Hematoxylin and eosin (H&E) stains are ubiquitous in pathology[1]. The H&E staining causes cell nuclei to turn blue and other tissue to turn pink[2]. Unfortunately, accurately identifying fine details of microanatomy on H&E-stained samples is challenging for those without pathology expertise[3], which makes large scale manual annotation of subtle structures costly and time intensive.

To counter the limitations of manual annotation on H&E, deep learning has been proposed as an alternate and automatic method for labeling microanatomy[4]. Deep learning algorithms are data hungry, meaning that performance generally improves as datasets increase in size[5]. Because manual annotation of cells is expensive and slow, the public release of large, labeled datasets in this space is important to facilitate the training of automatic cell identification algorithms. In 2022, the CoNIC Challenge released a dataset of colon H&E with six nucleus cell type annotations[3,6]. The challenge data was annotated using a complicated and repetitive approach that involved both automatic nucleus annotation and refinement based on feedback from trained pathologists[3,6]. Cell segmentation is an important and popular topic in digital pathology, with segmentations being useful in downstream applications[4,7–12]. The development of automatic and reliable nucleus classification algorithms for H&E slides would allow for more comprehensive large-scale cell tracking which promises better understanding of human physiology in both health and disease.

In contrast to H&E, multiplexed immunofluorescence (MxIF) imaging directly enables subclassification of cells. MxIF involves the staining and imaging of the same tissue multiple times via bleaching and re-staining[13]. When many stains are used in MxIF, a more detailed understanding of tissue structure can be attained than is available in H&E, because different stains can bind to different subsets of the tissue. When many stains are used on the same tissue, nuclei/cells can be classified based on which combinations of stains bind to each nucleus/cell.

The digital synthesis of unacquired stains is known as virtual staining[14]. Image slides can be virtually stained when the starting point is label-free or an acquired stain[15]. Taking an image of stained tissue and computationally generating a virtually stained image of the same tissue is known as stain-to-stain transformation[15]. Previous studies have used deep learning generative adversarial networks (GANs)[16] and conditional GANs[17] to synthesize virtual H&E images[18,19,20]. In this work, we refer to both synthetic and real data. To distinguish between acquired H&E and synthesized H&E, we henceforth refer to these as real H&E and virtual H&E, respectively.

MxIF is a specialized technology, making these images rare, whereas H&E is ubiquitous. It would be beneficial to derive the intricate microanatomical details present in MxIF images from H&E samples. Bridging the gap between MxIF and H&E can be formulated as a computational problem.

Co-leveraging H&E and MxIF information has been performed in several studies. Nadarajan et al. performed semantic segmentation of simple structures on real H&E using MxIF derived labels from the same tissue[21] with paired H&E and MxIF stains. In a follow-up paper, the same group used a conditional GAN to create virtual H&E from MxIF[2]. A semantic segmentation model was then trained on the virtual H&E with MxIF-derived labels to semantically segment 4 simple structures (all nuclei, cytoplasm, membranes, background) and evaluated on real H&E. Further work has been conducted in this area with Han et al. having designed a model that learned to classify 4 types of cells (ER+, PR+, HER2+, and Ki67+) from real H&E by leveraging paired MxIF information[22].

We investigate the more difficult task of learning to subclassify nuclei and cells into 14 categories on virtual H&E (Figure 1) and evaluate our models on real H&E from a public dataset. The contributions of this work are: 1) we demonstrate the degree with which 14 categories of nuclei/cells can be simultaneously learned on virtual H&E when using paired cell subclassification labels in a supervised training scheme, 2) we demonstrate the degree with which the virtual-trained model generalizes to real H&E, and 3) we are the first to automatically identify helper T and epithelial progenitor nuclei on H&E. This manuscript is a considerable extension of our prior work[23].

## 2    Methods

We studied an in-house dataset of MxIF images that were stained with 27 markers and used the inherent cell classification information to train a model to identify these cell categories on virtual H&E (Figure 2). We further evaluated the models on real H&E.

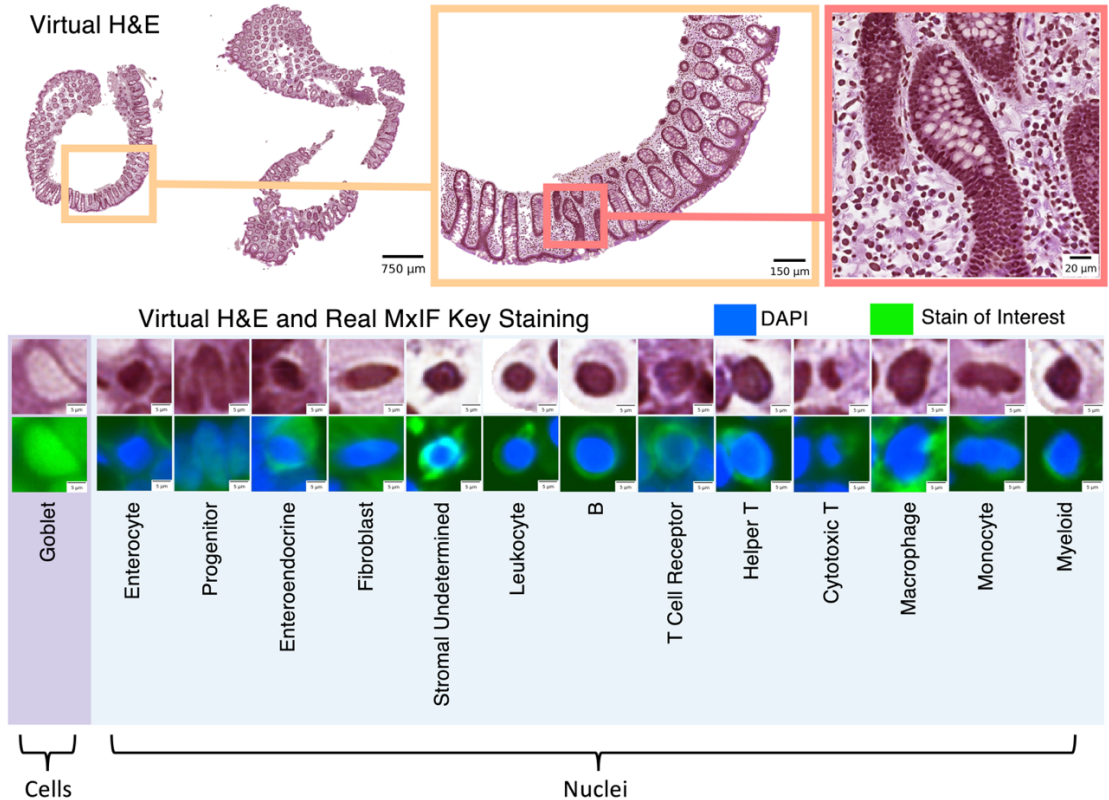

Figure 1. We leveraged inter-modality learning to investigate the identification of nuclei and cells on H&E staining that are traditionally viewed with specialized staining. The realistic quality of our virtual H&E holds at multiple scales (top section). Representative nuclei and cells from each of our 14 classes in both virtual H&E and MxIF illustrate intensity and morphological variation across cell types (lower section). Green is used to denote the MxIF stain of interest, which is a different stain for each of the 14 classes in this figure. While the signal to identify these classes of nuclei and cells is present in MxIF, the classes are more difficult to distinguish on virtual H&E.

In more detail, the MxIF images were style transferred to the H&E domain (§2.4). We used deep learning in a supervised training approach to learn to classify 14 types of nuclei and cells on virtual H&E in a multi-class classification approach (§2.5-2.7). The class label information was obtained for each cell or nucleus from the MxIF by using classification rules from biological domain knowledge about how combinations of markers bind to cells. Specifically, labels were generated for nucleus/cell classes via combinations of 17 out of 27 MxIF stains (§2.3 and Table 1). The virtual H&E was synthesized by using a CycleGAN[24] on all 27 MxIF stains.

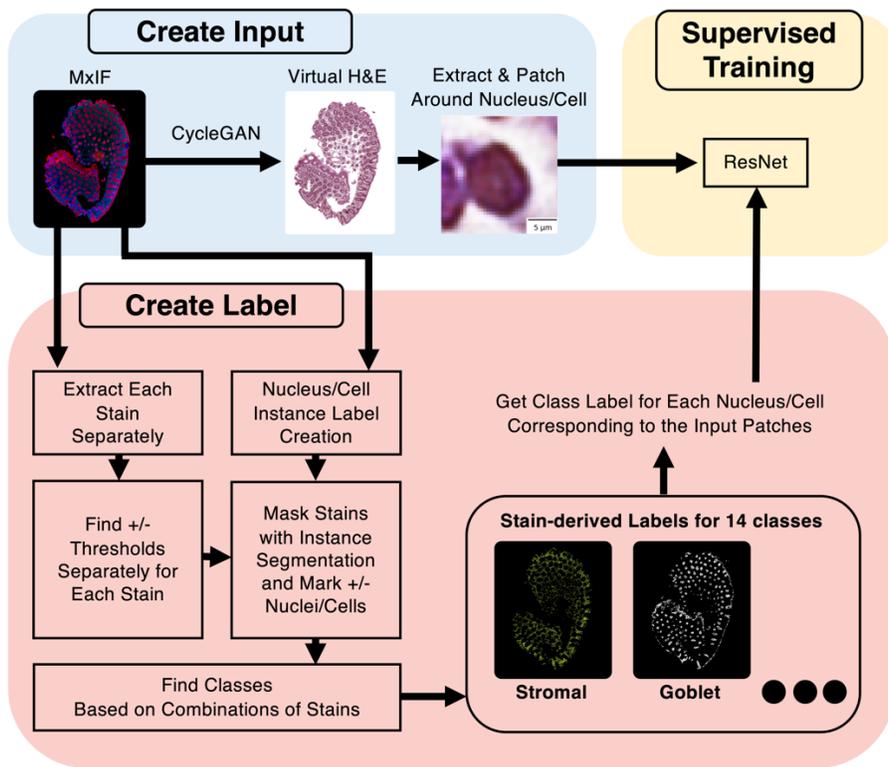

Figure 2. Our approach is similar to Han et al.,[22] with the additions of learning classification from virtual data and targeting the classification of 14 nucleus/cell classes rather than 4.

We trained a ResNet-18[25] on $41 \times 41$ pixel ($20.5 \times 20.5$ micron) image patches of virtual H&E with one nucleus or cell centered in each patch. The class label for each image patch corresponded to the center nucleus/cell and was derived from the stain combinations on the same nucleus or cell in the MxIF. We evaluated the models on both withheld virtual H&E, as well as public real H&E from a multi-site dataset.

*2.1 In-house MxIF Data*

Samples were studied in deidentified form from Vanderbilt University Medical Center under Institutional Review Board approval (IRB #191738 and #191777). The samples were labeled at the slide level by a pathologist as normal, quiescent, mild, moderate, or severe, with respect to Crohn's disease activity. The samples were formalin-fixed and paraffin-embedded. We used 28 whole slides imaged at 0.32 microns per pixel (14 from the ascending colon and 14 from the terminal ileum) from 20 patients. The distribution of slides per patient were as follows: 14 patients had one slide, five patients had two slides, and one patient had four slides. Slides were split at the patient level during training, validation, and testing, which we discuss in further detail in §2.8. We studied 17 out of 27 stain channels to annotate cell types on MxIF: NaKATPase, PanCK, Muc2, CgA, Vimentin, DAPI, SMA, Sox9, OLFM4, Lysozyme, CD45, CD20, CD68, CD11B, CD3d, CD8, and CD4 (see Table 1 for reasoning). Although we used 17 stains for annotation, we annotated 14 classes, because each of the 17 stains did not always directly map to a unique nucleus/cell type. These stains came from a subset of a previously described protocol[20].

*2.2 Public Real H&E Data*

The public real H&E data came from the CoNIC Challenge 2022 and is from normal colon tissue as well as colon tissue depicting cancer, dysplasia, and inflammation[6]. There were 4981 image patches of size 256×256 pixels which came from images with resolution ~0.5 microns per pixel. Nuclei in this dataset had labels for 6 cell types: epithelial, lymphocyte, plasma, eosinophil, neutrophil, and connective. The data came from multiple sites in the USA, England, and China[3]. Moreover, the data were aggregated from 5 source datasets: CRAG[26], GlaS[27], DigestPath, PanNuke[28], and CoNSeP[29].

Table 1. A sequential decision process was used for generating labels for the 14 nucleus/cell classes from MxIF. The bottom rows show final annotations for the 14 classes. These rules were used for labeling nucleus/cell instances.

| Step | Purpose | Stain Combinations |
|---|---|---|
| 1 | Group Epi+ instances | NaKATPase+ or PanCK+ or Muc2+ or CgA+ |
| 2 | Group Stroma+ instances | Vimentin+ or SMA+ |
| 3 | Exclude instances that are both Epi+ and Stroma+ | Exclude instances that are marked as (Epi+ and Stroma+) |
| 4 | Group Immune+ instances | Group instances that are CD45+ or CD20+ or CD68+ or CD11B+ or Lysozyme+ or CD3d+ or CD8+ or CD4+ |
| 5 | Remove immune conflicts for macrophage across all instances | Exclude instances where (CD68+ and CD3d+), (CD68+ and CD20+), (CD68+ and CD4+), (CD68+ and CD8+), or (CD68+ and CD11B+) |
| 6 | Remove immune conflicts for monocyte across all instances | Exclude instances where (CD11B+ and CD3d+), (CD11B+ and CD20+), (CD11B+ and CD4+), (CD11B+ and CD8+), or (CD11B+ and CD68+) |
| 7 | Remove immune conflicts for B cells across all instances | Exclude instances where (CD20+ and CD3d+), (CD20+ and CD4+), or (CD20+ and CD8+) |
| 8 | Remove conflicts for helper T and cytotoxic T across all instances | Exclude instances where (CD3d- and CD45- and CD4+), (CD3d- and CD45- and CD8+), or (CD4+ and CD8+) |
| 9 | Group Progenitor+ instances | Sox9+ or OLFM4+ |
| 10 | Exclude instances that are not in either the epithelium or stroma | Exclude instances that are (Epi- and Stroma-) |
| 11 | Remove conflicts for goblet cells across all instances | Exclude instances where (Muc2+ and Immune+), (Muc2+ and Progenitor+), or (Muc2+ and SMA+) |
| 12 | Remove conflicts for enteroendocrine across all instances | Exclude instances where (CgA+ and Immune+), (CgA+ and SMA+), (CgA+ and Progenitor+), or (CgA+ and Muc2+) |
| 13 | Remove conflicts for fibroblasts across all instances | Exclude instances that are (SMA+ and Immune+) |
| 14 | Remove conflicts for progenitors across all instances | Exclude instances where (Immune+ and Progenitor+) |
| 15 | Remove instances that are negative for all the large groupings | Exclude instances where (Epi- and Stroma- and Progenitor- and Immune-) |
| 16 | Remove any immune instances from Epi+ group | Exclude Epi+ instances where (Epi+ and Immune+) |
| 17 | Final annotation for goblet cells | Group Epi+ instances where (Muc2+ and Progenitor-) |
| 18 | Final annotation for enteroendocrine | Group Epi+ instances where (CgA+ and Progenitor-) |
| 19 | Final annotation for enterocyte | Group Epi+ instances where (CgA- and Progenitor- and Muc2-) |
| 20 | Group fibroblasts/stromal (undetermined) | Group instances that are Stroma+ and Immune- |
| 21 | Final annotation for fibroblasts | Group fibroblast/stromal (undetermined) where (SMA+ and Progenitor-) |
| 22 | Final annotation for stromal (undetermined) | Group fibroblast/stromal (undetermined) where (SMA- and Progenitor-) |
| 23 | Final annotation for myeloid | Group Immune+ instances that are ((Lysozyme+ and CD68- and CD11B- and Progenitor- and CD20-) and CD3d- and CD8- and CD4-) |
| 24 | Final annotation for helper T | Group Immune+ instances where (CD4+ and Progenitor-) |
| 25 | Final annotation for cytotoxic T | Group Immune+ instances where (CD8+ and Progenitor-) |
| 26 | Final annotation for T cell receptor | Group Immune+ instances where (CD3d+ and CD4- and CD8-) |
| 27 | Final annotation for monocyte | Group Immune+ instances where (CD11b+ and CD3d- and Progenitor- and CD4- and CD8-) |
| 28 | Final annotation for macrophage | Group Immune+ instances where (CD68+ and CD3d- and Progenitor- and CD4- and CD8-) |
| 29 | Final annotation for B cell | Group Immune+ instances where (CD20+ and CD68- and CD3d- and Progenitor- and CD4- and CD8-) |
| 30 | Final annotation for leukocyte | Group Immune+ instances where (CD45+ and CD20- and CD68- and CD3d- and Progenitor- and CD4- and CD8- and CD11B- and Lysozyme-) |
| 31 | Final annotation for progenitor | Group all instances that are Progenitor+ |

*2.3 Label Generation Leveraging MxIF*

Generating classification labels for nuclei and cells required identifying whether a stain bound or did not bind to cells at the whole slide image level. To determine stain binding required picking a threshold for positive and negative nuclei and cells, for each stain. In this work, a senior digital pathology researcher manually selected and applied stain-wise thresholds. These thresholds were determined separately by stain channel for each MxIF whole slide image.

To determine the location of each nucleus/cell, we obtained an instance segmentation by performing inference with the pretrained DeepCell Mesmer model[30]. We passed the Mesmer model a single grayscale channel image as input. This single channel input was the sum of the MxIF DAPI and Muc2 channels. Merging the channels through addition was reasonable, since DAPI identifies nuclei, and Muc2 identifies the goblets associated with goblet cells. Each nucleus or cell was then categorized as positive or negative for each stain type by computing its mean stain intensity and applying the manual thresholds. We assigned each instance a single class label based on a series of biological rules (Table 1). After label generation, we had 14 classes, 13 of which were nucleus classes: enteroendocrine, enterocyte, epithelial progenitor, fibroblast, stromal (undetermined), monocyte, macrophage, helper T, cytotoxic T, T cell receptor, B, myeloid, and leukocyte, and one of which was a cell class: goblet.

We differentiate goblet cells as not being nuclei, because they were identified via MUC2 positive regions, which are the goblets from goblet cells, rather than the nuclei of the goblet cells. For progenitors, since 96% were epithelial positive in our dataset, we refer to progenitors and epithelial progenitors interchangeably throughout this manuscript. Additionally, the nucleus class stromal (undetermined) refers to cells in the stroma that were not classified further into any specific subclasses.

## 2.4 Virtual H&E via Style Transfer

To train the models for classification of nuclei/cells on virtual H&E, we needed image-label pairs of virtual H&E (from MxIF) and cell labels (from MxIF). The virtual H&E was inferred from all 27 MxIF stains using a pretrained CycleGAN-based model that performed style transfer. This model was trained on in-house MxIF and in-house real H&E. The architecture and training strategy (named "Proposed-(8)") were described in detail in a previous work from our team[20].

## 2.5 Nucleus/Cell Identification and Classification Approach

To train a cell classification model that could later be applied to unlabeled H&E data, we needed to have a pipeline that consisted of two major components: nucleus/cell localization and nucleus/cell classification. In this pursuit, our pipeline consisted of an instance segmentation model and a separate classification model (on image patches surrounding each nucleus/cell). When the identification and classification were used together and evaluated, the classification performance was based on matched nuclei. Predicted nuclei were matched to labeled nuclei if the intersection-over-union (IoU) was greater than 0.5, as in the Hover-Net paper[29].

## 2.6 Nucleus/Cell Identification Model & Training Approach

The instance segmentation model we used on virtual H&E was a Hover-Net[29]. We trained the Hover-Nets on virtual H&E input with matching instance segmentation labels, predicted from the pretrained DeepCell Mesmer model on MxIF (the data was downsampled via cubic interpolation to 0.5 microns per pixel so that the Hover-Nets would work at the same resolution as our target real H&E data[3,6]). The training strategy that we selected was the default from the Hover-Net public github repository: https://github.com/vqdang/hover_net.

*2.7 Nucleus/Cell Classification Model & Training Approach*

We selected ResNet-18[25] as our classification model. The model was initialized from the PyTorch 1.12.1 default ResNet-18[25] pretrained on ImageNet[31]. To remove negative effects from class imbalance and batch size differences between training and testing, we replaced each batch normalization[32] layer with instance normalization[33]. The instance normalization layers were implemented with PyTorch using default parameters.

The ResNet-18 was trained for classification on image patches of virtual H&E. The patches were resampled via cubic interpolation to a standard H&E resolution of 0.5 microns per pixel. Our patch extraction strategy involved selecting patches of size 41 x 41 pixels centered on a nucleus or cell. All patches were individually intensity normalized between 0 and 1. A single class label was assigned to each patch, corresponding to the center nucleus or cell (Figure 1). To address heavy class imbalance during training, at each batch, we ensured that each example had an equal likelihood of coming from any of the 14 classes. The model was trained for 20,000 steps using a batch size of 256, the Adam optimizer[34], a learning rate of 0.001, cross-entropy loss, and the one cycle learning rate scheduler[35].

All of the classification code was implemented in Python 3.8 using PyTorch 1.12.1 and Torchvision 0.13.1. Additionally, all classification training and inference was performed using an Nvidia RTX A6000. The code is available at the repository at this link: github.com/MASILab/nucleus_and_cell_classification_on_he

*2.8 Cross-validation*

We performed five-fold cross-validation separately on both the instance segmentation and classification models on the virtual H&E. To avoid data contamination, the training, validation, and testing data were split at the patient level (Figure 3). To maintain consistency across folds, we specified that in each fold, the training data contained 12 patients, validation contained 4 patients, and testing contained 4 patients. To reduce bias, we always included data from both the ascending colon and terminal ileum, healthy and diseased, in each training, validation, and testing set.

As mentioned in §2.3, on our virtual H&E we created labels for goblet cells, which we identified by finding the Muc2+ goblets associated with the cells, rather than by finding their nuclei. Goblets look very different from cell nuclei on H&E (Figure 1). On the real H&E data that we selected for external validation, there were no labels on the goblets associated with goblet cells, because this data came from the CoNIC Challenge, which only dealt with nuclei. So, we needed to train two instance segmentation models (Hover-Nets): the first to segment nuclei and goblets so that we could evaluate our approach on our virtual H&E, and the second to only segment nuclei, so that we could evaluate on the real H&E from the CoNIC Challenge.

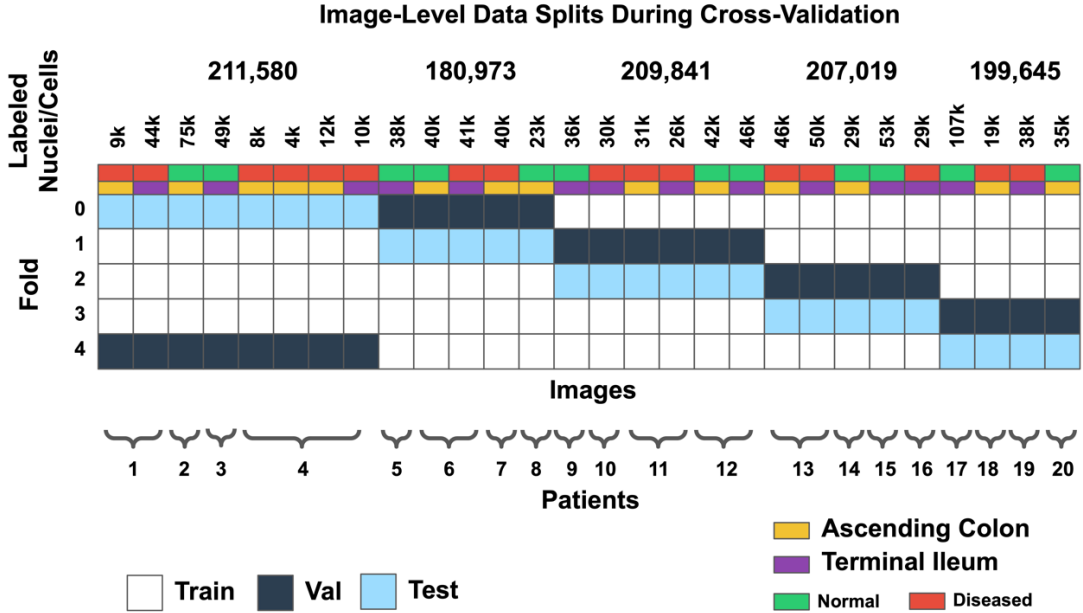

Figure 3. To balance the quality of the models trained on each fold of data, we split training, validation, and testing data at the patient level (each column is an image). Additionally, each training, validation, and testing fold contained tissue without Crohn's disease activity (normal) and tissue with Crohn's disease activity (diseased) from the ascending colon and terminal ileum.

We selected weights for evaluation based on the step with the lowest validation loss for each fold. Evaluation was performed on the segmentation and classification models both separately and combined.

### 2.7 External Validation on Real H&E

Our entire cross-validation approach was performed on virtual H&E. To test how well our virtual-trained models generalized to unseen real H&E data, we used the publicly available CoNIC dataset[6]. This dataset contained 4981 image patches from five source datasets. To reduce the effects of the domain shift between the stain color of our virtual H&E and the real H&E data, we trained a separate CycleGAN to make the real H&E staining more similar to our virtual H&E. The style transferred patches of real H&E were then used for external validation of the segmentation and classification approach.

## 3   Results

*3.1 Results for Nucleus/Cell Instance Segmentation*

The instance segmentation was evaluated with precision and recall (Figure 4). In this work, for evaluation of the instance segmentation performance, we considered a predicted nucleus/cell to be a true positive if any pixels overlapped with an instance label. If there was no label overlap on a prediction, then the nucleus/cell was a false positive. If a label had no overlapping prediction, then the nucleus/cell was a false negative. Defining true positive, false positive, and false negative at the nucleus/cell level, rather than pixel level, was necessary to account for the differences in the label sets between our virtual H&E and real H&E. More specifically, the instance segmentation labels on our virtual H&E (inferred by the pretrained DeepCell Mesmer model) were not as closely cropped to the boundary of each nucleus/cell as the labels on the real H&E from the CoNIC dataset.

On the virtual H&E, precision and recall were high whether or not the Hover-Nets were made to include goblet cells for instance segmentation. On the real H&E, the recall was high across source datasets. However, the precision on real H&E varied across source datasets and had a drop in performance compared to the virtual H&E due to a larger number of false positives.

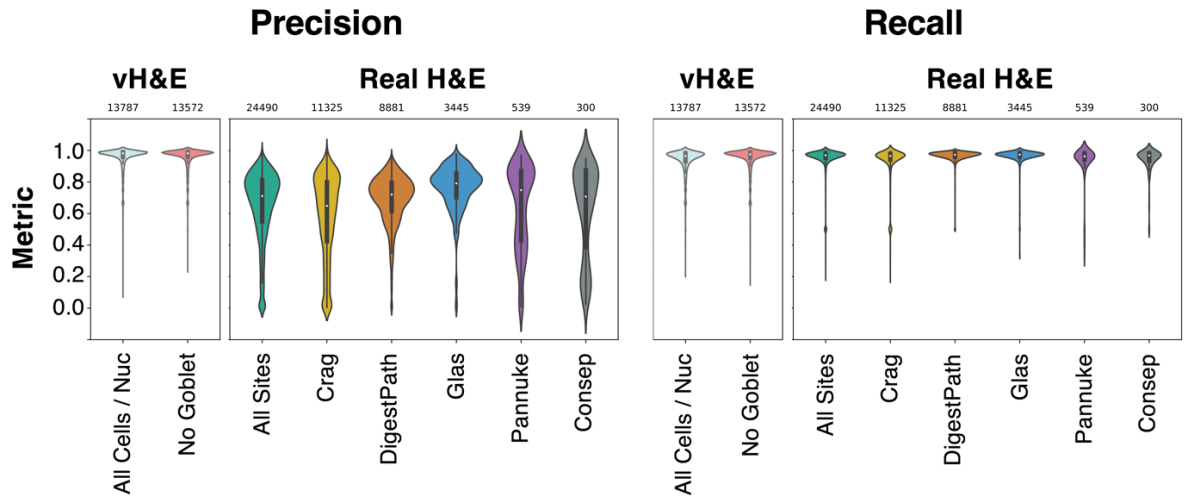

Figure 4. Instance segmentation was stable on recall across virtual H&E and real H&E, but less stable on precision. Lower precision on the external real H&E means that there were more false positives, according to the label set.

In Figure 5, we further investigated the lower precision for instance segmentation on the real H&E through visualization. The model sometimes confused darker colored style transferred structures with nuclei. Additionally, the labels on the real H&E did not always contain every nucleus. Both factors contributed to a higher number of false positives which caused the reduced precision.

### 3.2 Results for Virtual H&E Nucleus/Cell Classification

On virtual H&E, the classification accuracy of the ResNet was usable for a subset of classes (Figure 6). These classes were helper T, macrophage, enterocyte, epithelial progenitor, enteroendocrine and fibroblast nuclei, as well as goblet cells. We show that this classification was stable when the classification was performed using the ground truth centroid information, as well as when the centroids were predicted with the Hover-Nets.

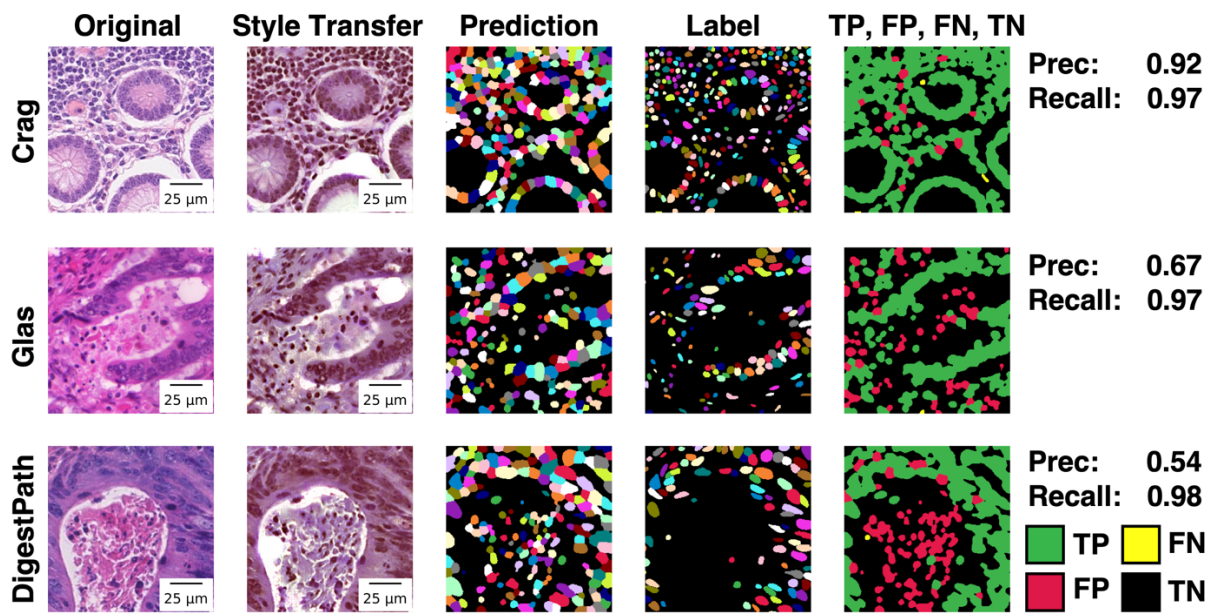

Figure 5. To counteract domain shift in H&E staining, instance segmentation was performed on a style transferred version of the real H&E. Lower precision on instance segmentation occurred when 1) the labels were not exhaustive and 2) style transfer made tissue darker and present more like nuclei. Note that the colors for nuclei are not expected to match between the prediction and label columns—these colors were used to visually separate instances, but do not have any classification meaning.

Looking in more detail at classification performance on virtual H&E, we computed the positive predictive value (PPV), negative predictive value (NPV), and prevalence (Figure 7). When prevalence is low, we expect PPV to be low and NPV to be high. Likewise, when prevalence is high, we expect PPV to be high and NPV to be low. A cutoff for reliable classification could be selected based on PPV (Figure 7 at 0.3). The classes above this threshold are helper T, enterocyte, epithelial progenitor, fibroblast, and stromal (undetermined) nuclei, as well as goblet cells. When considering prevalence, we note PPV is high for helper T and epithelial progenitor nuclei, and NPV is high for goblet cells and enterocyte nuclei. These results are relatively stable when using the ground truth centroids or predicted centroids. Qualitative results of nucleus/cell classification are reasonable and detailed in Figure 8.

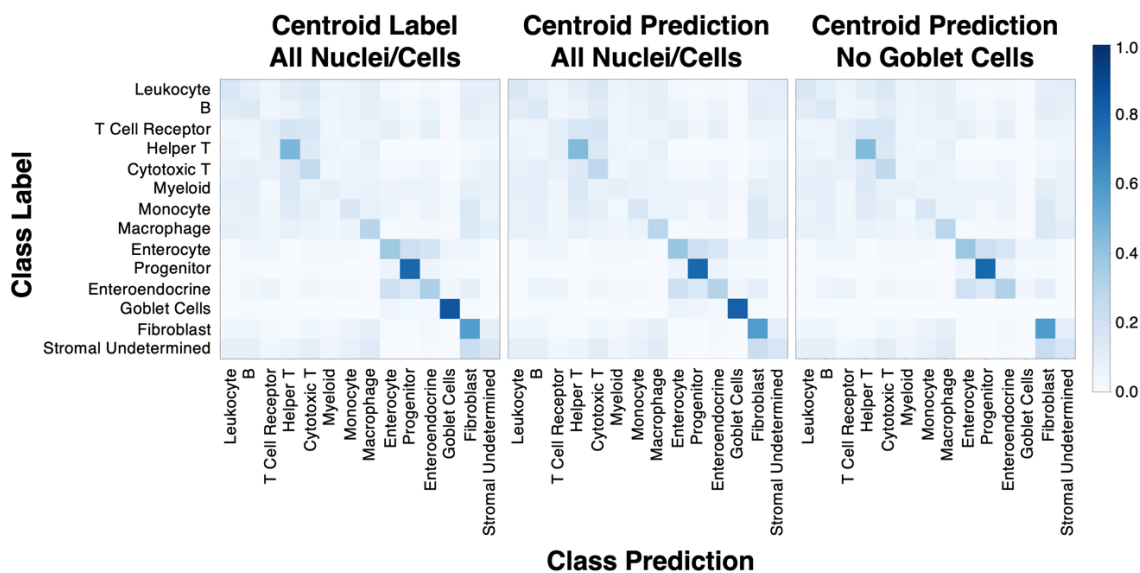

Figure 6. While not all classes were learned from virtual H&E (shown for 5-fold cross-validation), learning behavior can be seen for helper T, macrophage, enterocyte, progenitor, enteroendocrine, and fibroblast nuclei, as well as goblet cells. The classification model's ability to learn MxIF label information (derived from 17 stain channels) on virtual H&E (3 RGB channels) implies that there is signal present in our virtual H&E to learn some of our fine-grained classification subtypes. From left to right, we show classification accuracy is stable when using label centroids, predicted centroids for all nuclei/cell types, and predicted centroids for all categories excluding goblet cells.

### 3.3 Results for Real H&E Nucleus Classification

We evaluated our virtual-trained models on real H&E, for five of the six classes that showed learning behavior on PPV and NPV (helper T, enterocyte, epithelial progenitor, fibroblast, and stromal undetermined). The sixth class that was not transferred was goblet cells. In our virtual H&E, we identified goblet cells by the goblet, not the nucleus. Because there were no appropriate corresponding labels in the real H&E data (only nuclei were labeled), we excluded goblet cells from this evaluation. We matched our selected virtual H&E classes to the closest matching and available parent classes on the real H&E. The matching scheme between the virtual H&E nucleus classes to real H&E nucleus classes was as follows: helper T to lymphocyte, enterocyte to epithelial, epithelial progenitor to epithelial, fibroblast to connective, and stromal (undetermined) to connective. Because we matched to parent classes, we could not compute PPV and NPV directly, and so instead computed bounds on PPV and NPV.

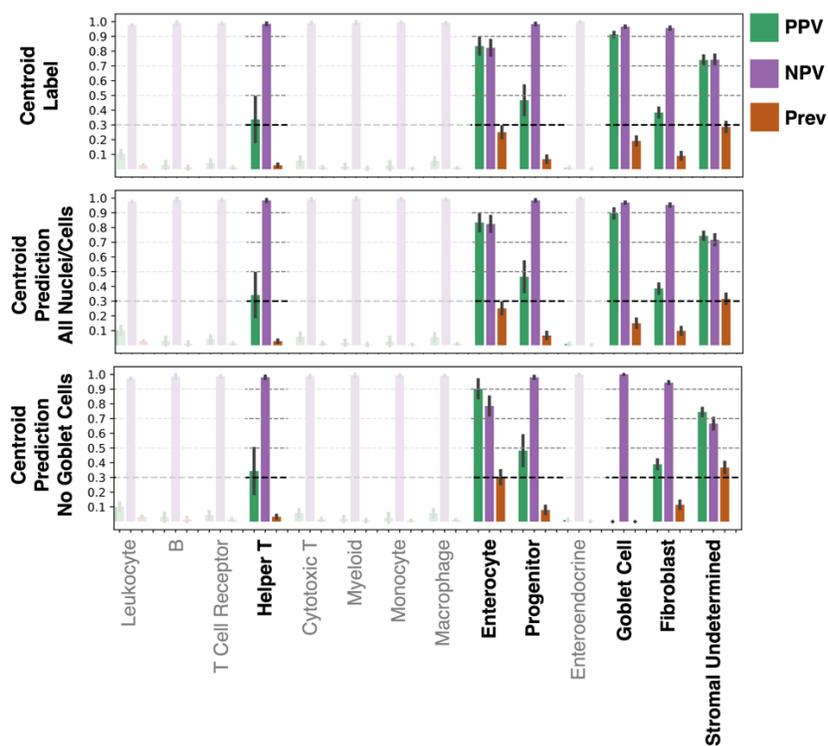

Figure 7. Virtual H&E metrics of PPV, NPV, and prevalence are shown highlighting which classes can be reasonably identified. These metrics give more insight into the usefulness of model predictions than accuracy shown in a confusion matrix. Helper T, enterocyte, epithelial progenitor, fibroblast, and stromal (undetermined) nuclei, as well as goblet cells, show reasonable learning. Additionally, performance is relatively stable when comparing classification using ground truth centroids, predicted centroids (all nuclei/cells), or predicted centroids (all except goblet).

The information learned on virtual H&E can transfer to an external testing set of real H&E (Figure 9). Looking in more detail, we compared performance on different datasets/sites of real H&E by using prevalence-normalized PPV (PPV divided by prevalence). We found prevalence-normalized PPV showed instability across datasets/sites, despite the use of style transfer to address differences in staining (Figure 10).

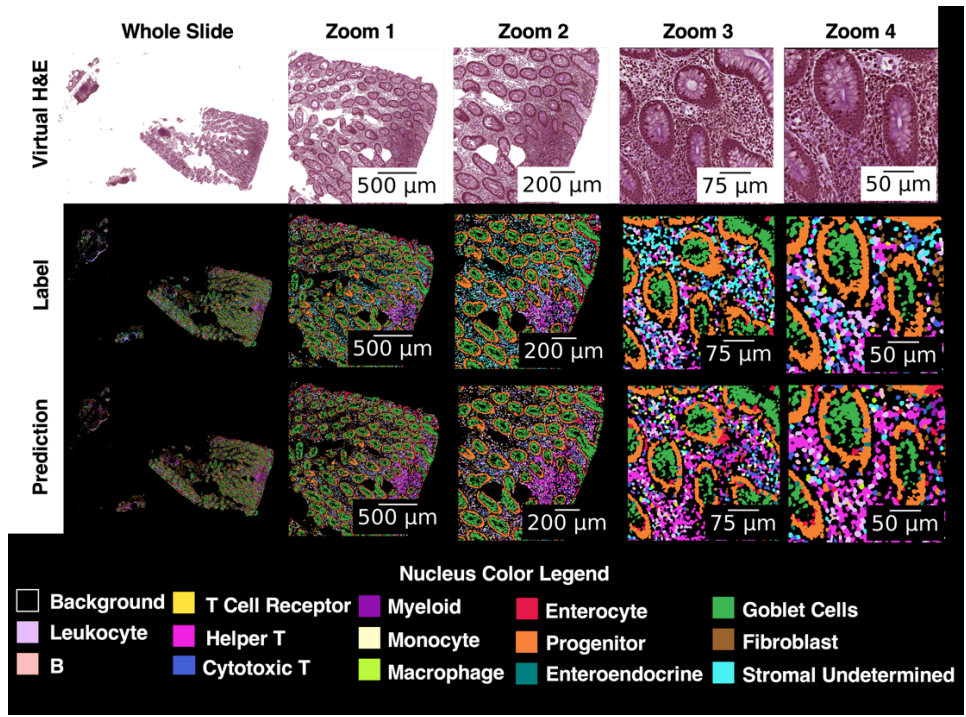

Figure 8. Classification is visually consistent at multiple zoomed scales for enterocyte and progenitor nuclei, as well as goblet cells. Stromal (undetermined) nuclei are often incorrectly classified as a variety of immune cells. Nuclei in an area with immune activity are often correctly identified as helper T cells, though the predictions do include a non-trivial number of false positives for helper T. Crypts, visible here as oval structures at zoomed scales mostly outlined by collections of progenitor (orange) nuclei, are expected to be comprised of epithelial cells. However, progenitors around the crypts in the label set are not surprising, since 96% of the progenitors in this dataset are positive for stains that identify epithelium.

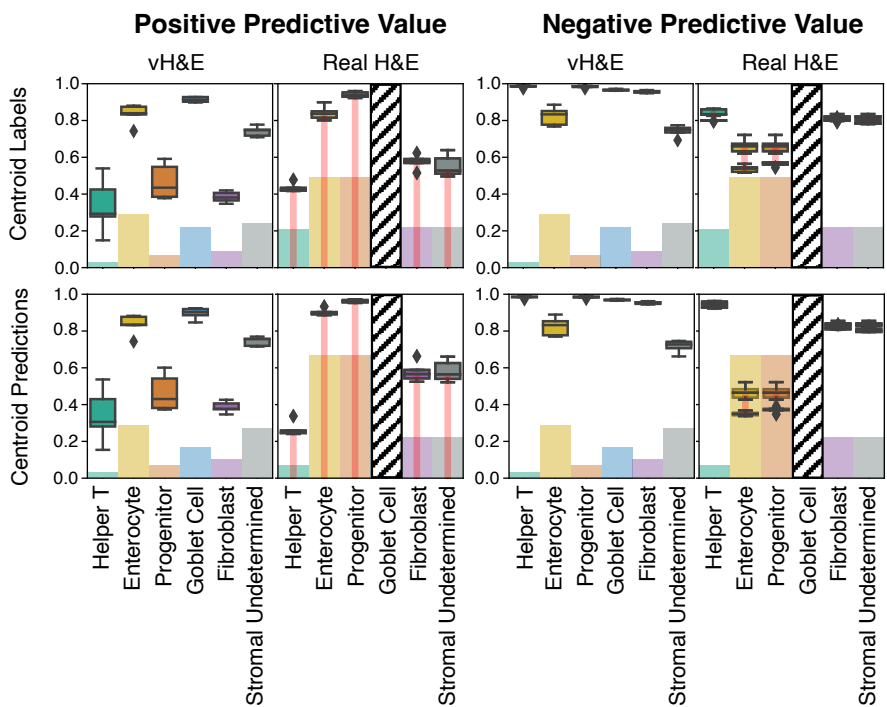

Figure 9. Nucleus classification learned on virtual H&E transfers reasonably well to the real H&E external testing data, despite it being diverse and coming from multiple source datasets. Boxplots for positive predictive value and negative predictive value are shown for both virtual H&E and all source datasets from the real H&E on a subset of cell types. The bar plots denote class prevalence for virtual H&E and the closest matching class prevalence on the real H&E. A thin red connecting bar is used on real H&E to denote that the metrics available are upper bounds on PPV and upper and lower bounds on NPV. Because of the lack of real H&E labels at the granularity of our cells of interest, our NPV and PPV are bounded metrics based on the available parent class labels—this means that we can only compute an upper bound for PPV, which is optimistic, and an upper and lower bound for NPV which allows for an estimate of NPV. Goblet cell performance is not shown for real H&E, because the goblets were not labeled in the real H&E testing data and so there was no appropriate available matching label.

## 4    Discussion & Conclusion

In this work, we trained a nucleus/cell subclassification model for H&E by leveraging inter-modality learning to train models on virtual H&E with MxIF label information for 14 classes.

Identification and classification of nuclei and cells was reasonably learned on virtual H&E for these classes: helper T, epithelial progenitor, enterocyte, fibroblast, and stromal (undetermined) nuclei, as well as goblet cells. Validation was performed on real H&E for helper T, epithelial progenitor, enterocyte, fibroblast, and stromal (undetermined) nuclei.

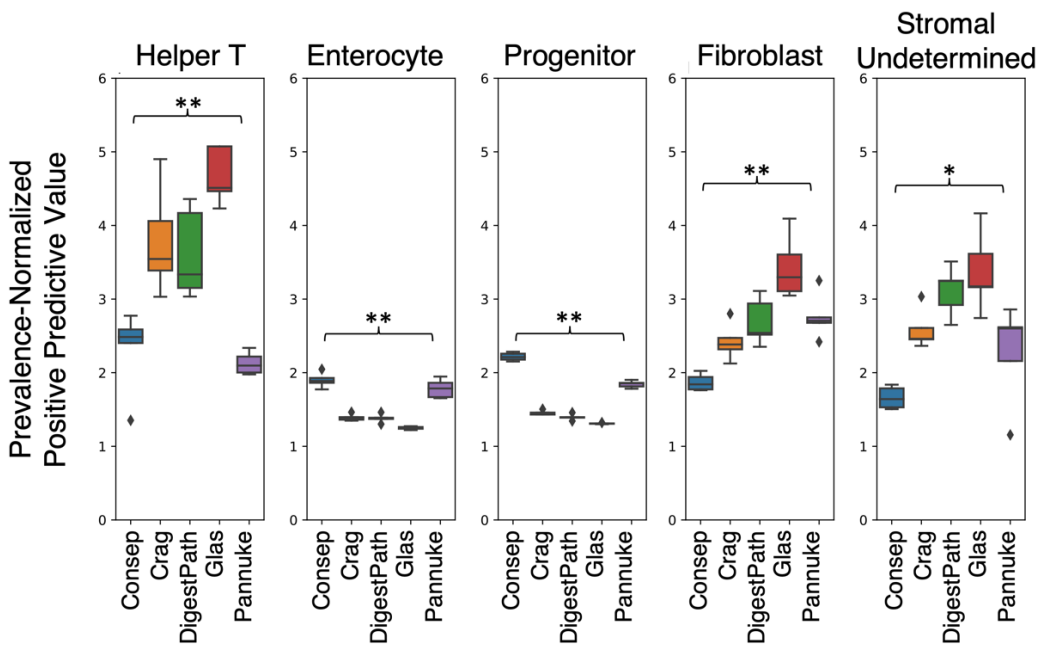

Figure 10. To assess the stability of our virtual-trained cell identification and classification pipeline on real H&E, we evaluated our approach across the source datasets within the real H&E dataset (Consep, Crag, DigestPath, Glas, Pannuke). Because the real H&E does not have labels to the granularity of our shown subclasses (helper T, enterocyte, epithelial progenitor, fibroblast, stromal undetermined), the PPV that we computed is an upper bound based on the matching available parent class labels. To compare the upper bound PPV across datasets for each cell type, we normalized by class prevalence (upper bound PPV divided by prevalence), which varies across source datasets. For each cell type shown, we computed the Friedman test to assess for differences across datasets. Values of $p < 0.05$ are denoted with * and $p < 0.001$ with **. Statistical significance for helper T, enterocyte, epithelial progenitor, fibroblast, and stromal (undetermined) nuclei backs up what is visually evidenced by the boxplots— performance on the PPV upper bound varies across datasets.

While it is feasible to create a large number of labels for nucleus/cell subtypes from MxIF, many of these labels are not easily learned on paired H&E-like data. This is not surprising, as specialized stains are commonly used to isolate many of the nuclei/cells we attempted to learn to identify, such as helper T cells. However, for helper T, epithelial progenitor, enterocyte, fibroblast, stromal (undetermined) nuclei, and goblet cells, there is some learnable information in our virtual H&E.

Due to the lack of previous works performing multi-class nucleus classification on H&E using MxIF stain label information, it is difficult to compare the performance of our model to the literature. In the similar work from Han et al., 4 types of nuclei (ER+, PR+, HER2+, and Ki67+) were identified from real H&E information with AUCs >= 0.75 using MxIF stain label information[22]. These markers were used to assess breast cancer samples and were not present in our marker panel. Additionally, their metric was for binary classification rather than multi-class classification, so a direct quantitative comparison is not reasonable.

In Figure 10, we observed intra-class variability on the prevalence-normalized PPV upper bound across datasets/sites in the real H&E testing data. While we used a CycleGAN for style transfer to correct for H&E stain variation across sites, perhaps more sophisticated harmonization approaches will be needed in the future to account for inter-site variability.

This work is limited by a lack of labels on the real H&E testing data at the same level of cell type granularity as our virtual H&E labels. This required us to compute bounds on metrics rather than the metrics themselves. A more convincing approach for learning MxIF label information on H&E stains would require having H&E staining and MxIF on the same histological samples (same cut), which would eliminate the need for virtual H&E.

Classification of nuclei and cells on H&E is promising for helper T, epithelial progenitor, enterocyte, fibroblast, stromal (undetermined) nuclei, as well as goblet cells. The ability to discern nucleus/cell subtypes based on shape and H&E staining is an exciting prospect in computational pathology. We have released code related to this manuscript at this online repository: github.com/MASILab/nucleus_and_cell_classification_on_he

**Disclosures**

K.S.L. is an hourly scientific consultant for Etiome, Inc.

**Code and Data Availability**

The H&E data from the CoNIC Challenge is publicly available here: https://conic-challenge.grand-challenge.org/. Code from this work is available here: https://github.com/MASILab/nucleus_and_cell_classification_on_he.

*Acknowledgments*

This publication is part of the Gut Cell Atlas Crohn's Disease Consortium funded by The Leona M. and Harry B. Helmsley Charitable Trust and is supported by a grant from Helmsley to Vanderbilt University www.helmsleytrust.org/gut-cell-atlas/. This research was supported by NSF CAREER 1452485, NSF 2040462, and in part using the resources of the Advanced Computing Center for Research and Education (ACCRE) at Vanderbilt University, Nashville, TN. This work was supported by Integrated Training in Engineering and Diabetes, grant number T32 DK101003. This material is partially supported by the National Science Foundation Graduate Research Fellowship under Grant No. DGE-1746891. This project was supported in part by the National Center for Research Resources, Grant UL1 RR024975-01, and is now at the National Center for Advancing Translational Sciences, Grant 2 UL1 TR000445-06, Department of Veterans Affairs grants I01BX004366, I01CX002171, and I01CX002473. We would like to acknowledge the VUMC Digestive Disease Research Center supported by NIH grant P30DK058404. This work is supported by NIH grants T32GM007347, R01DK135597, R01DK103831, R01DK128200, R01EB017230. We extend gratitude to NVIDIA for their support by means of the NVIDIA hardware grant. We have used AI as a tool in the creation of this content, however, the foundational ideas, underlying concepts, and original gist stem directly from the personal insights, creativity, and intellectual effort of the author(s). The use of generative AI serves to enhance and support the author's original contributions by assisting in the ideation, drafting, and refinement processes. All AI-assisted content has been carefully reviewed, edited, and approved by the author(s) to ensure it aligns with the intended message, values, and creativity of the work. The Vanderbilt Institute for Clinical and Translational Research (VICTR) is funded by the National Center for Advancing Translational Sciences (NCATS) Clinical Translational Science Award (CTSA) Program, Award

**Biography**

Lucas W. Remedios is a PhD student in Computer Science at Vanderbilt University. His research focuses on artificial intelligence in medical imaging domains. He is advised by Dr. Bennett Landman.

Other author biographies are not  available.